# Enhanced Performance of Highly Activated Carbon and Surface-Treated Porous Organic Polymers as Physical Adsorbents for Chemical Warfare Agents

Sanghyeon Park[1], Yuseung Hong[2], Hyunseo Choi[2*]

[1]Materials & Energy Technology Center, [2*]Chem-Bio Technology Center, Agency for Defense Development, Daejeon 34186, Republic of Korea

**Abstract**

The use of chemical warfare agents (CWAs) in modern warfare cannot be disregarded due to their ease of use and potential for large-scale incapacitation. An effective countermeasure involves the physical adsorption of these agents, preventing their entry through the respiratory tract by non-specific adsorption. In this study, we investigate the physical interaction between potential adsorbents and model gases mimicking CWAs, thereby identifying sufficient conditions for higher physisorption performance. Our findings reveal that the physisorption capacity is highly sensitive to the surface properties of the adsorbents, with uniform development of micropores, rather than solely high surface area, emerging as a critical factor. Additionally, we identified the potential of porous organic polymers as promising alternatives to conventional activated carbon-based adsorbents. Through a facile introduction of polar sulfone functional groups on the polymer surface, we demonstrated that these polar surface polymers exhibit physical adsorption capabilities for formaldehyde under ambient conditions comparable to high-performance activated carbons. Notably, the superior activated carbon possessed a high BET surface area of 2400 m²/g and an exceptionally uniform micropore structure with an average pore size of approximately 11 Å. This research paves the way for designing adsorbents with high physical adsorption capacities tailored for CWAs protection, offering a significant advancement in developing next-generation protective materials.

keywords: Chemical Warfare Agent, Adsorption, Activate Carbon, Porous Organic Polymers

## 1. Introduction

Around May 2024, North Korea's spraying of filth balloons against the South occurred, increasing tensions between the two Koreas. [1] In addition, in July, a fallen filth balloon led to a fire. Until now, only simple garbage piles such as manure, manure, glass pieces, and detonators have been found in the filth balloon. However, if an unknown chemical agent is contained in the balloon and delivered to South Korea, it will cause even more damage. This study was proposed to increase the capability to respond to the risks of terrorism and warfare situations, including the unknown Chemical Warfare Agents (CWAs), which are being heightened by the various delivering means of attack.

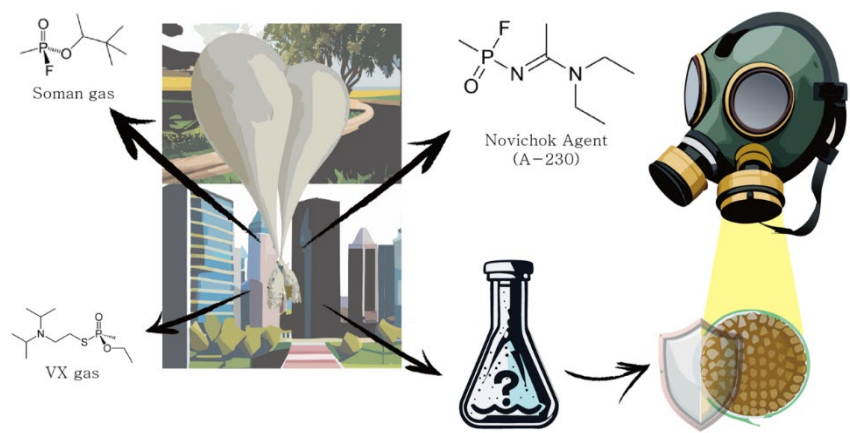

**Fig 1**. Scenario of a situation in which Enemy state sends unknown chemical warfare agents (Created By Adobe illustrator 2024, Font: Adobe Myungjo Std m)

The principle of military gas masks was used to develop an inhibitor that can effectively cope with CWAs. The current purification tank of military gas masks is based on activated carbon, one of the porous materials, and the protection principle is to adsorb the CWAs in the respiratory tract to the activated carbon so that it does not reach the respiratory tract [2]. Adsorption is divided into physical and chemical adsorption according to the interaction between the adsorbent molecule and the adsorbent, each accompanied by a van der Waals interaction and a direct chemical bond. Physical adsorption (physisorption) and chemical adsorption (chemisorption) usually occur inevitably together. [3] At this time, physisorption can adsorb almost all the agents regardless of the type of CWAs, and the adsorption amount is larger due to multilayer adsorption. On the other hand, chemisorption has a specificity between the agent and the impregnated material (metal such as copper) on the surface of the adsorbent. In the early stages of using CWA, it is not possible to grasp the chemical weapons used by the enemy, and considering this, it is the performance of physical adsorption that has an absolute effect on survival from CWA. Therefore, in this study, to improve the performance of the gas mask purification tank, it is intended to improve the physical adsorption capacity through material development, and ultimately save more lives.

This study does not limit the adsorbent to activated carbon alone, but carefully examines the possibility of other porous materials as physical adsorbents to improve the physical adsorption capacity. These porous materials are largely divided into two types: activated carbon and porous organic polymers (POPs). In addition, an adsorption isotherm analysis and a possibility review were conducted on an imidazole salt-based metal organic framework (MOF) material called ZIF-8, but it was found that the adsorption performance was low under low-pressure conditions (~1atm) in the actual chemical atmosphere due to the lack of open metal site [4]. Therefore, in this study, the performance was improved and potential was examined for two porous materials, activated carbon, which is used as a physical adsorbent for purification tanks, and POPs. The physical adsorption ability of the adsorbent is determined by the specific surface area of the material, the development of uniform micropores, and the affinity between the CWAs and the surface of the adsorbent. This study, which deals with activated carbon and POPs materials, basically aimed to find a material with the above three factors evenly and to synthesize an activated carbon material with a BET specific surface area ($S_{BET}$) that is more than twice as high as domestic and foreign commercial activated carbon (DARCO, A.C., 6006K, 3M.), which is a comparative group in the case of activated carbon. In addition, in the case of POPs, since various functional groups can be synthesized and attached through modification [5], a specific functional group (sulfone functional group) was introduced on the material surface to increase the affinity with the polar CWA molecule. Through this, it aimed to confirm the clearly improved physical adsorption capacity compared to POPs without modification.

In this study, the pores of substances were analyzed in more detail to investigate the physical adsorption process in more detail, because in nano-porous materials, the physical adsorption process is dominated by the interaction between fluid-fluid and fluid-solid and the interaction within the fluid's pore. [6] Therefore, the analysis process is as follows. Model Gas ($N_2$, $CO_2$, $CH_2O$) was used for safety in the adsorption performance analysis, and ASAP, FESEM, and XRD equipment were used. Through $N_2$ isotherm analysis conducted in 77 K, texture properties of adsorbents are scrutinized, including the specific surface area and pore size distribution, etc. After that, $CO_2$ isotherm analysis was conducted, and $Ph_6M_4Cs_{11}$ showed the highest adsorption amount, and there was no significant difference between the physisorption capacity of PP-2-S and PP-2. In the last $CH_2O$ isotherm analysis, unlike the previous $CO_2$ experimental results, the amazing adsorption performance of PP-S-2 comparable to that of $Ph_6M_4Cs_{11}$ was confirmed. Thereby we revealed the potential of porous organic polymers as a CWAs physisorption substrate, in terms of the ease of applying well-working polar functional groups. FESEM and XRD equipment were used to analyze the morphology of substances, and the Hobart-Kawazoe (HK) method and the Barrett-Joyner-Halenda (BJH) method were used to understand the morphology of substances in more detail through pore distribution.

## 2. Method

### 2-1. Model Gas Selection

In this study, three model gases (nitrogen, carbon dioxide, and formaldehyde) were selected to effectively simulate the interactions between the synthesized adsorbent and CWAs, with the aim of evaluating the protective performance of the adsorbent. The interactions between chemical agents and adsorbents can be broadly categorized into four primary types: (1) π-π stacking between the π bonds of the adsorbent's hexagonal ring or phosphorus-oxygen (P=O) functional group and those of the adsorbate; (2) C-H-π interactions between the hydrogen in C-H bonds of adsorbate and the π bonds on the surface of the adsorbent; (3) dipole-dipole interactions between polar functional groups of adsorbent and adsorbate; and (4) coordination bonding between the adsorbate, acting as a ligand, and the metal sites on the adsorbent's surface. [7]

The activated carbon adsorbent considered in this study consists of a continuous nonpolar structure made up of hexagonal carbon rings, which are expected to interact with the nonpolar regions of CWAs. The three selected model gases, all possessing multiple bonds, are hypothesized to engage in π-π interactions with the adsorbent's surface, which is rich in π bonds. Additionally, among the model gases, $N_2$ and $CO_2$ are nonpolar, while $CH_2O$ is a polar molecule. Therefore, $CH_2O$ is expected to exhibit dipole-dipole interactions and van der Waals forces with the polar functional groups of the adsorbent.

For nitrogen gas, can be used to measure the physical properties of the synthesized adsorbent, such as specific surface area and pore size distribution. [8] The adsorption performance of various synthesized porous adsorbents was analyzed using nitrogen gas at 77K, allowing the determination of their physical properties. Adsorption isotherms at room temperature were analyzed using carbon dioxide and formaldehyde to measure the adsorption capacity of each synthesized porous adsorbent. By comparing the adsorption performance of the nonpolar carbon dioxide and the polar formaldehyde, we examined the differences in adsorption behavior based on the presence of molecular polarity.

**2-2. Porous Adsorbent Synthesis**

**2-2-1. Activated Carbon ($Ph_6M_4Cs_{11}$)**

2,4-Dihydroxybenzoic acid (Ph(s)) (0.600 g, 3.85 mmol), melamine (M(s)) (0.323 g, 2.56 mmol), and CsOH(aq) (1.2 ml) were added to a 70 ml vial. Subsequently, 10.95 ml of DI water was added, and the vial was sealed and stirred at 450 rpm for 1 hour at 60°C to achieve a homogeneous solution. 0.972 ml of formaldehyde solution was added, and stirring was continued under the same conditions for an additional hour, during which a dark reddish-brown color was observed in the solution. The solution was then dried at 120°C for 24 hours, allowed to cool to room temperature, and further dried under vacuum to obtain the melamine-phenolic resin, which serves as the precursor for activated carbon.[9]

The sample was transferred to the tubular furnace, and to suppress potential side reactions such as combustion during the synthesis process, an adequate argon atmosphere was established within the furnace. The sample was then carbonized at 900°C for 2 hours. As a result, the sample, which expanded in volume due to the activation of micropores, was stirred in a 5 wt% HCl(aq) for 12 hours. Subsequently, the sample was washed with DI water until the filtrate became clear, and then vacuum filtration was performed using a Nylon 6/6 membrane with a pore size of 0.2 micrometers. After filtration, impurities such as inorganic salts remaining in the sample were removed through washing. The sample was then dried under vacuum at 180°C for 12 hours to synthesize the activated carbon.[9]

**2-2-2. Porous organic polymers**

· **PP-2**

Anhydrous $FeCl_3$ (1.622 g) was added to a solution of dichloro-para-xylene (para-DCX) (10.0 mmol, 1.751 g) in 65 ml of dichloroethane (DCE). The reaction was conducted via Friedel-Crafts alkylation, with stirring at 45°C for 1 hour, followed by stirring at 80°C for an additional hour, resulting in the synthesis of the porous organic polymer, PP-2. To remove unreacted residues and obtain high-purity PP-2 from the synthesized sample, filtration was performed using a Nylon 6/6 membrane (pore size 0.2 micrometers) and washing was carried out with tetrahydrofuran and methanol until the filtrate became clear. The sample (PP-2) was then dried in a vacuum oven at 60°C for 20 hours. [5]

· PP-2-S

To investigate the extent of interaction with model gases based on the presence of polar functional groups on the surface of the porous organic polymer, a sulfonic group was added to PP-2 to synthesize the functionalized PP-2-S. The overall process is similar to that of PP-2 synthesis, but dichloromethane (DCM, 23 ml) was used as the solvent, and chlorosulfonic acid was added using a dropping funnel to 0.4 g of pre-synthesized PP-2. Specifically, 23 ml of DCM and 0.4 g of PP-2 were added to the reaction pot and stirred until the solution was homogeneous. Chlorosulfonic acid (2.8 ml) and 10 ml of DCM were added dropwise via the dropping funnel. The reaction mixture was then stirred in an ice bath at 0°C for 1 hour, followed by exposure to room temperature and stirring for an additional 24 hours. Over time, the color of the solution was observed to turn black. The reaction solution was washed with DCM and methanol, and vacuum filtration was performed using a hydrophobic PTFE filter until the filtrate was clear. The sample was then dried in a vacuum oven at 60°C for 20 hours to obtain the functionalized porous organic polymer, PP-2-S. [5]

## 3. Results and Discussion

Highly-activated carbon ($Ph_6M_4Cs_{11}$)[9] was synthesized and its surface features and adsorption capacity are examined, as well as prepared commercial activated carbon (DARCO) for thorough comparison with $Ph_6M_4Cs_{11}$. A representative and potential porous organic polymer, namely PP-2 [5] was also synthesized and sulfonated PP-2 (PP-2-S) was newly synthesized to consider the future use of these polymers for CWAs adsorbents, where the activated carbons are widely and primarily used in both military and industry fields. [10] Since our study mainly focuses on physisorption adsorbents, requiring no singularity with specific CWAs, we used $CO_2$ and $CH_2O$ as model gases to investigate inter-pi and inter-dipole-moment interactions between adsorbents and adsorbates, rather than using limited and tricky real CWAs (e.g. Novichok, Sarin, and VX).

### 3-1. Materials Characterization

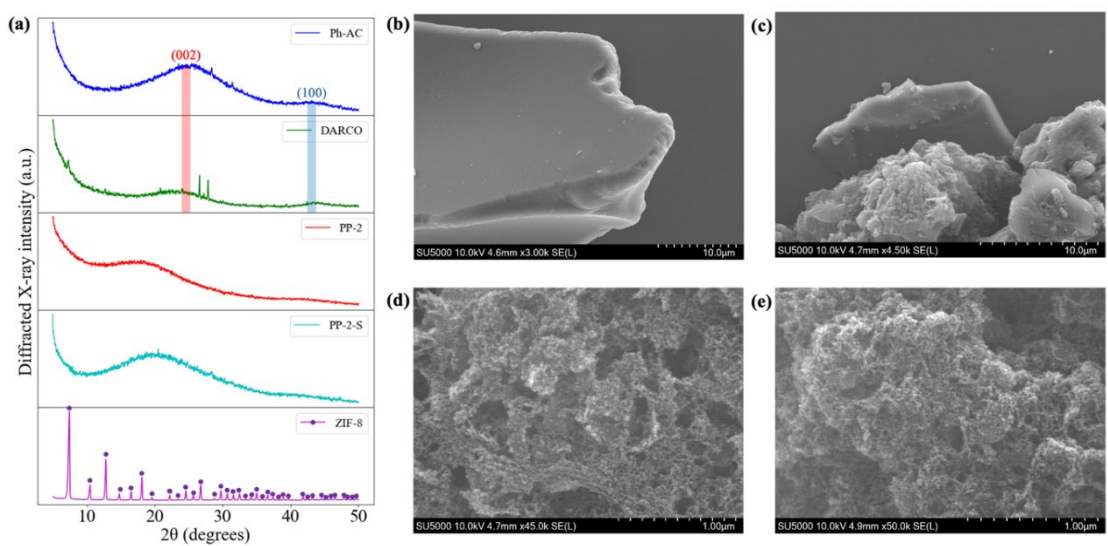

**Fig 2**. (a) Powder X-ray diffraction (PXRD) patterns of activated carbons ($Ph_6M_4Cs_{11}$ and DARCO), porous organic polymers (PP-2 and PP-2-S) (Created by Python Idle, Fonts: sans-serif), and a representative metal organic framework (ZIF-8) with Cu Kα radiation. (b-e) Field Emission Scanning Electron Microscopy (FE-SEM) images for (b) $Ph_6M_4Cs_{11}$ (as-synthesized), (c) commercial DARCO, and as-synthesized PP-2 and PP-2-S ((d) and (e), respectively).

Structural and morphological features of porous physical adsorbents are determined through powder X-ray diffraction (PXRD) patterns and Field-Emission scanning electron microscopy (FESEM). Fig. 2(a) displays PXRD patterns of as synthesized $Ph_6M_4Cs_{11}$, PP-2, and PP-2-S, as well as commercial activated carbon (DARCO) and ZIF-8, a representative metal organic framework. [11] Peak broadening of upper 4 materials clearly reflects the amorphous nature of activated carbons and porous organic polymers, in contrast to sharp and narrow peaks of as-synthesized ZIF-8, which correspond to the known international center of diffraction database's (ICDD)ZIF-8 data. For two activated carbons ($Ph_6M_4Cs_{11}$ and DARCO), two common broad peaks around 25° and 43° are the typical peaks of the (002) and (100) planes of the carbon, respectively. [12] It means that partial organization has occurred during the pyrolysis process, so some graphitic phases are locally evolved. [13] Therefore, it might be inferred that local ordering of polymer (PP-2 and PP-2-S) has also been occurred during the synthesis, considering the unknown broad bands around 20°.

It should be also noted that the actual intensity (counts per second, cps) of all of the four amorphous adsorbents are about 3-order lower than that of ZIF-8 due to the lack of crystallinity. Therefore, tiny effect of impurity (e.g. quartz) or sample holder could be easily detected as unknown two sharp and narrow peaks around 26° and 28° of the PXRD pattern of DARCO. Fig. 2. (b)-(e) are FESEM images for $Ph_6M_4Cs_{11}$, AC DARCO, PP-2, and PP-2-S, respectively, visualizing surface morphologies of each of the adsorbents. Clearly, surface morphologies are classified into two categories; hard monolith for activated carbons, and xerogel for porous organic polymers, consistent with previous experimental studies. [5], [9] In addition, it can be seen that the morphology of DARCO is not complete hard monolith, supporting that some impurities are incorporated, reasonably consistent with the PXRD pattern.

## 3-2. Materials Texture with $N_2$ Isotherms

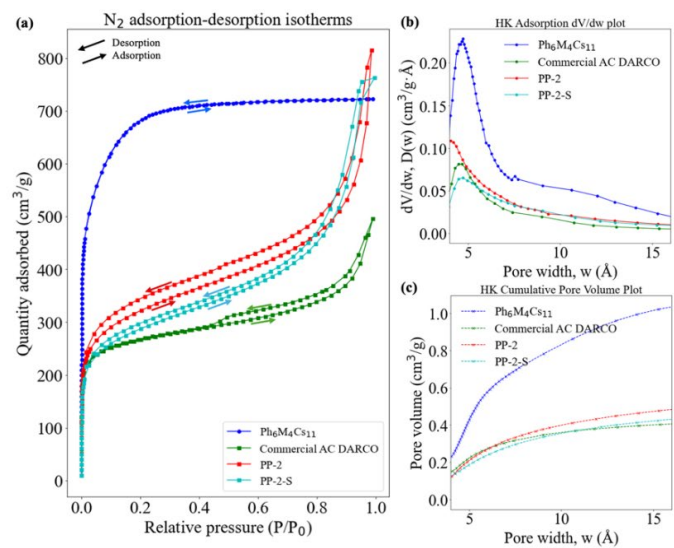

**Fig 3**. (a) $N_2$ adsorption-desorption isotherms of $Ph_6M_4Cs_{11}$, DARCO, PP-2, and PP-2-S. Lower and upper bound of each of the hysteresis are adsorption and desorption isotherms, respectively. P0 is equal to 1 atm. (b) Differential and (c) cumulative pore volume plot obtained through the Horvath-Kawazoe (HK) method. (Created by Python Idle, Fonts: sans-serif)

Subsequently, $N_2$ adsorption-desorption isotherms at 77 K (the boiling point of nitrogen) have been obtained, in order to get porosity-related information. Fig. 3(a) displays $N_2$ isotherms of $Ph_6M_4Cs_{11}$, DARCO, PP-2, and PP-2-S. All of 4 $N_2$ isotherms are classified into the IUPAC adsorption-desorption isotherm classification, thereby the texture information (e.g. Pore size distribution, pore geometry, and connectivity) is qualitatively available. To begin with, it can be seen that the $N_2$ isotherm of $Ph_6M_4Cs_{11}$ perfectly follows type I isotherm (namely Langmuir isotherm) according to the IUPAC classification. Therefore, it is clear that $Ph_6M_4Cs_{11}$ is microporous with well-defined uniform micropores where the pore width is less than 2 nm. Plateau of this isotherm indicates that the accessible micropore volume predominantly governs the $N_2$ uptake, and the internal surface area doesn't participate in the adsorption process. [14] In addition, the absence of hysteresis also notifies that the adsorbed $N_2$ molecules are under similar interactions with $Ph_6M_4Cs_{11}$, also indicating the uniformity of predominant micropores.

In contrast, another activated carbon (DARCO), easily available in the market, shows type IV isotherm with hysteresis loop, indicating a broad distribution of pores in both microporous and mesoporous ranges. Simultaneously, synthesized two porous organic polymers (PP-2, and PP-2-S) also demonstrate type IV isotherm like DARCO. These hysteresis loops are associated with capillary condensation taking place in mesopores. [15] Besides the hysteresis loop, unlike the type I isotherms, these three isotherms have a common point that the amount of nitrogen absorption increases linearly without saturation, and then increases exponentially as the pressure increases. Therefore, it could be inferred that monolayer adsorption has already been completed at low pressure then multilayer adsorption has occurred, qualitatively suggesting that the specific surface areas of DARCO, PP-2, and PP-2-S are fairly less than that of $Ph_6M_4Cs_{11}$.

Among three types IV isotherms, it should be noted that the two isotherms of porous organic polymers differ from the DARCO isotherm in that the hysteresis loop starts immediately in the low-pressure section where the pressure is almost zero. We believe that this is due to differences in the distribution of pore sizes, and in the case of DARCO, the hysteresis loop started later under higher pressure conditions because it has a relatively narrow distribution compared to porous organic polymers. In this regard, the distribution of pore size is correlated with the hysteresis because of the fact that gas molecules fill smaller pores first, and the larger pores are filled subsequently due to the difference in the thermodynamic site preference.

In order to quantitatively investigate the distribution of the pore width ranging from the micro- and meso- sizes (0-500 Å), we have conducted both the Horvath-Kawazoe (HK) and the Barrett-Joyner-Halenda (BJH) method [16][17]. Fig. 3(b) and (c) show differential (D(w)=dV/dw) and cumulative pore volume (V) plot, respectively, and those two are obtained from the HK method, using slit pore model, and providing the pore size distribution information about micropores (0-20 Å). According to those plots, it is clear that highly activated carbon, $Ph_6M_4Cs_{11}$ show narrow width distribution in micropore window, considering that the $Ph_6M_4Cs_{11}$ show sharper peak and steeper slope than those of the other adsorbents, in differential- and cumulative- plot, respectively. Bae et. al. insist that this narrow distribution is arising from the strategic preparation of precursor before carbonization process; melamine-rich precursor, where the uniformly-existing melamine-like-structure undergoes pyrolysis, resulting in the formation of uniformly-sized micropores. [9] Although those plots from the HK method provide unclear information about which adsorbent poses a narrower distribution between DARCO and PP-2, PP-2-S is seemed to have the broadest distribution in the micropore range, among 4 adsorbents.

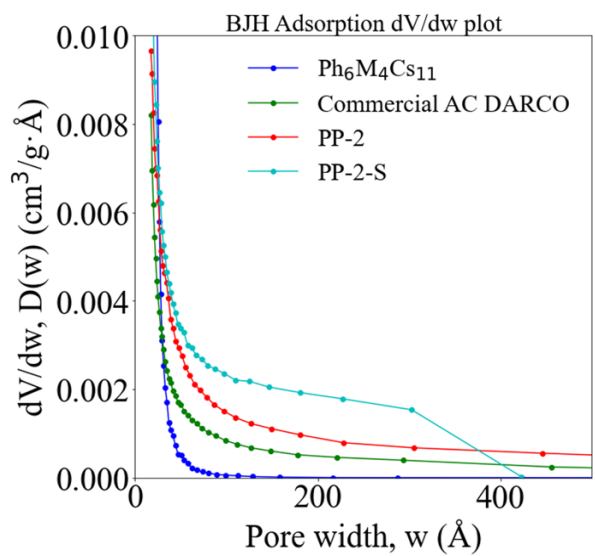

**Fig 4**. Barrett-Joyner-Halenda (BJH) Adsorption differential pore volume plot (D(w)) in the range from 0 to 500 Å.
(Created by Python Idle, Fonts: sans-serif)

The uniformity of the pore size beyond the micropore range is also investigated via the BJH method, as demonstrated in Fig. 4. Surprisingly, the distribution results provided by the BJH method clearly showed superiority order in whether uniformly-sized pores were formed in the four adsorbents: $Ph_6M_4Cs_{11}$

showed the most uniform pore size distribution, followed by DARCO, PP-2, and PP-2-S in order. Note that it is clear that PP-2-S has the broadest pore size distribution, and we believe that the existing micropores in PP-2 have participated in surface reaction related to sulfonation, resulting in the collapse of micropores (smaller pores) into mesopores (larger ones), thereby the pore distribution is further broadened in both micro- and meso-range.

|  | $Ph_6M_4Cs_{11}$ | Darco | PP-2 | PP-2-S |
|---|---|---|---|---|
| Averaged micro-pore width (Å) | 8.33 | 7.07 | 6.27 | 8.95 |
| Averaged meso-pore width (Å) | 31.23 | 148.91 | 157.73 | 155.32 |
| **Averaged pore width (Å)** | 11.40 | 69.38 | 65.11 | 105.01 |

Table 1. Average micropore and mesopore width obtained from integration of distribution plot. Average pore width regarding the distribution in 0-500 Å is also tabulated.

The average pore width ($\bar{w}$) is sometimes described as a pore width indicated by the distribution's peak location in previous studies, however, we have directly integrated the distribution from the two methods to get the average pore size where the broadness of the distribution is reflected on, with the following equation:

$$\bar{w} = \frac{\int_a^b wD(w)dw}{\int_a^b D(w)dw}$$

Equation 1. The integration method for obtaining averaged pore width where D(w) is the distribution of the pore width as a function of pore width (w). Consequently, averaged micropore width ([a, b]=[0, 20]) and mesopore width ([a, b]=[20, 500]) are tabulated together with the average pore width ([a, b]=[0, 500]), as shown in the table. 1. Firstly, the average micropore width ($\bar{w}_{micro}$) is about 1.43 times increased during the sulfonation, evidently indicating the surface etching has occurred as previously mentioned. Although the average mesopore width ($\bar{w}_{meso}$) of PP-2-S is slightly little than that of PP-2, it also means that smaller pores mainly participate in surface sulfonation, while larger pores do not tend to participate in. In addition, the average pore size ($\bar{w}$) of the $Ph_6M_4Cs_{11}$ is 11.4 Å, indicating that there is almost no mesopores as also shown in the BJH pore width distribution. In contrast, $\bar{w}$ of the PP-2-S is 105.0 Å, 1.61 times increased than that of PP-2 (65.1 Å) shows the disadvantages of introduction of sulfone functional groups into the surface, in terms of the texture profile.

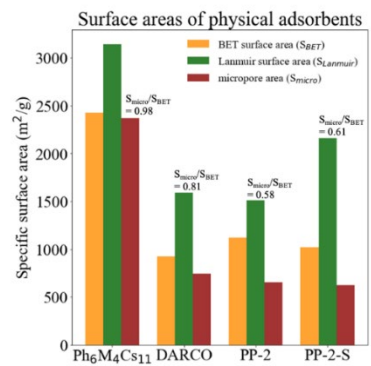

**Fig 5**. A comparison between three specific surface areas (including $S_{BET}$, $S_{Lanmuir}$, and $S_{micro}$) of 4 adsorbents: $Ph_6M_4Cs_{11}$, Darco, PP-2, and PP-2-S. (Created by Python Idle, Fonts: sans-serif)

|  | Ph$_6$M$_4$Cs$_{11}$ | DARCO | 3M-6006K | 3M-3311K-100 | PP-2 | PP-2-S | ZIF-8 |
|---|---|---|---|---|---|---|---|
| S$_{BET}$ (m$^2$/g) | 2,426 | 922 | 984 | 1,455 | 1,122 | 1,020 | 1,314 |
| S$_{Lanmuir}$ (m$^2$/g) | 3,141 | 1,593 | 1,412 | 1,899 | 1,511 | 2,161 | 1,830 |
| S$_{micro}$ (m$^2$/g) | 2,372 | 742 | 910 | 1,418 | 653 | 627 | 1,296 |
| S$_{micro}$/ S$_{BET}$) | 0.98 | 0.81 | 0.93 | 0.97 | 0.58 | 0.62 | 0.99 |

Table 2. Summary for specific surface areas of as-synthesized (**Ph$_6$M$_4$Cs$_{11}$**, PP-2, PP-2-S, ZIF-8) and commercially used (DARCO, 3M-6006K, 3M-3311K-100) physical adsorbents.

The specific surface area can also be determined through the inert gas (N$_2$ at the 77 K) adsorption isotherm. Fig. 5. and table. 2. illustrates the specific surface areas of various physical adsorbents, including Ph$_6$M$_4$Cs$_{11}$, DARCO, PP-2, PP-2-S, ZIF-8, and additionally investigated commercial activated carbons (3M-6006K and -3311K-100). The adsorbent Ph$_6$M$_4$Cs$_{11}$ exhibits the highest values across the all three surface areas, Brunauer-Emmett-Teller (BET) specific surface area (S$_{BET}$) [18] slightly above 2400 m$^2$/g, Langmuir specific surface area (S$_{Langmuir}$) around 3000 m$^2$/g, and specific surface area of micropores (S$_{micro}$) slightly below 2400 m$^2$/g, indicating that 98% of the BET surface area is due to micropores. Additionally, it should be noted that, according to the table. 2., two porous organic polymers (PP-2 and PP-2-S) have less micropore contribution onto the BET surface area than activated carbons, including Ph$_6$M$_4$Cs$_{11}$, DARCO, and additionally investigated commercial activated carbons. In other words, activated carbon seems to develop micropores better than porous organic polymers, and we believe that this is the fundamental difference of the material itself stemming from that porous organic polymers do not undergo carbonization like activated carbon and thus have a wider pore size distribution. [19]

### 3-3. CO$_2$ Adsorption Capacities

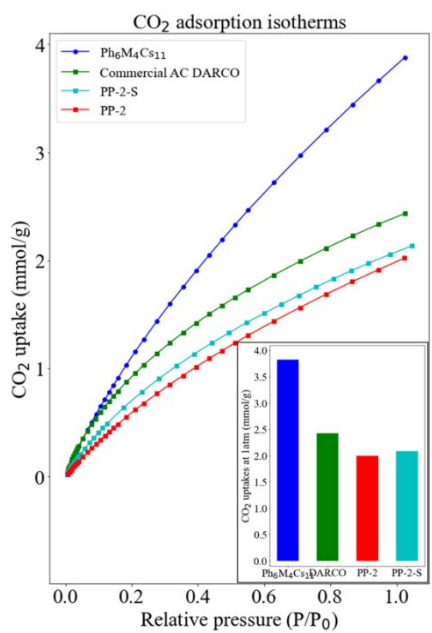

**Fig 6**. CO$_2$ adsorption isotherms obtained at room temperature. CO$_2$ uptakes (mmol/g) at 1 atm are also plotted as below bar graph. (Created by Python Idle, Fonts: sans-serif)

The evaluation of $CO_2$ adsorption performance was conducted. $CO_2$, which is a non-polar molecule and has two double bonds, can simulate the pi interaction between actual CWAs and adsorbents for physisorption. All $CO_2$ analyses were conducted at room temperature (298 K), and the $CO_2$ adsorption isotherms of the four types of porous materials are shown in Fig. 6. Based on the adsorption performance at 1 atm of AC DARCO (2.43 mmol/g), a commercial activated carbon, activated carbon $Ph_6M_4Cs_{11}$ shows about 1.5 times better performance (3.82 mmol/g), and porous organic polymers, including PP-2 and PP-2-S, have adsorption capacities of 2.00 mmol/g and 2.09 mmol/g, respectively. In the formaldehyde adsorption performance analysis described later, the dipole interaction between adsorbate and adsorbent significantly governs the adsorption performance, indicating that the sulfone functional group introduced on the surface of PP-2 does not play a significant role in the increase of adsorption capacity for non-polar $CO_2$ molecules. Instead, $CO_2$ adsorption capacity is rather governed by the texture profile, especially the microporosity, as the order of adsorption performance of the four adsorbents is the same as that of $S_{micro}/S_{BET}$ (Fig.5). Park et. al. pointed out that adsorbents with lower $S_{BET}$ (1787 m$^2$/g) but higher microporosity (~92%) with appropriately sized micropore shows better $CO_2$ physisorption performance than another adsorbent (2047 m$^2$/g and ~74%), supporting our perspective that the uniform micropore growth is essential in efficiently capturing the non-polar molecular weapons. [20] Therefore, based on this point of view, we strongly believe that the physisorption capacity through pi-interaction of actual CWAs and physical adsorbents is dominantly influenced by surface properties, especially the formation of uniform micropores.

### 3-4. $CH_2O$ Adsorption Capacities

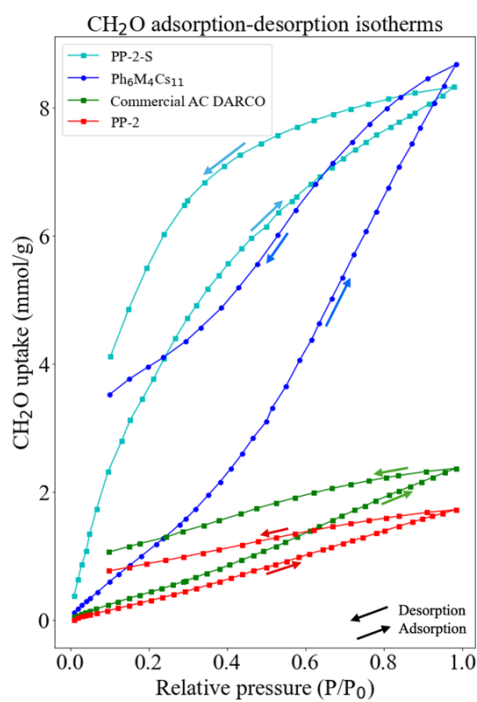

**Fig 7**. $CH_2O$ adsorption-desorption isotherms obtained at room temperature. (Created by Python Idle, Fonts: sans-serif)

Subsequently, formaldehyde ($CH_2O$) adsorption performance evaluation is followed. $CH_2O$ is a model gas that can simulate the dipole-dipole interaction between an actual CWAs and an adsorbent, and $CH_2O$ isotherms obtained at 298 K can show the results of surface modification by introducing a polar functional group to PP-2, as illustrated in Fig. 7. Firstly, PP-2-S shows superior adsorption capacity for $CH_2O$ molecule, overwhelming that of $Ph_6M_4Cs_{11}$ in low pressure range, and being comparable to that of $Ph_6M_4Cs_{11}$ in atmospheric pressure. This superior performance indicates that polar functional group on surface is efficiently working for dipole-dipole interaction with $CH_2O$ molecule. Simultaneously, $Ph_6M_4Cs_{11}$ boasts outstanding physisorption performance compared to both AC DARCO and PP-2, stemming from that the physisorption capacity from the dipole-induced-dipole interaction between adsorbate and adsorbent is also believed to be largely influenced by the texture property of physisorption adsorbent. In addition, it should be worth mentioning that isotherm curve of PP-2-S represents the saturating curve as pressure increases, unlike to isotherms of other adsorbents. This implies that $CH_2O$ molecules are likely to adsorb onto the surface of PP-2-S as monolayer, further supporting that the introduction of polar functional group is successfully and intentionally working where the $CH_2O$ molecules forms energetically stable interactions with the functional group on the PP-2-S surface. Isotherms of all four adsorbents represent hysteresis phenomena arising from the capillary condensation, which is the formation of less strongly bound between $CH_2O$ layers and capillary-adsorbed $CH_2O$ than much strongly bound $CH_2O$ monolayer. Considering that similar hysteresis phenomena occur in the water adsorption-desorption isotherm [21][22], we believe that this large hysteresis of $CH_2O$ isotherm stems from especially the polar nature of $CH_2O$ adsorbate.

Consequently, introduction of functional group with polarity was significantly helpful for enhancing physisorption capacity via the dipole-dipole interaction, when the target adsorbate molecule is polar. Therefore, in terms of ease in functionalization, porous organic polymers are promising as the physical adsorbent for CWAs with molecular polarity. Additionally, even though the adsorbent is a polar molecule, superior texture profile is still significant for the high adsorption capacity through the dipole-induced-dipole interaction between adsorbent and adsorbate. In this study, we have aimed to figure out both the significance of texture profile and the potential of porous organic polymers as the physisorption adsorbent for chemical agents. In this study, however, thermal and/or mechanical stability has not been scrutinized since those were out of the scope. Herein, we believe that the thermal and mechanical stability of porous organic polymers could be appropriately tailored for use as molecular weapon adsorbent, by choosing a suitable polymer backbone (e.g. rigid aromatic polymer), so such investigation will be further studied in future work. [23] In addition, the composite of activated carbons with superb texture properties and functionalized-and-robust porous organic polymers will be further studied for the use of physisorption adsorbent for unknown chemical agents, thereby paving the way to successfully protect lives from deadly chemical warfare.

## 4. Conclusion

Once chemical warfare occurs, it takes time to determine the exact species of used CWAs. Therefore, physisorption capacity should be crucially regarded, as physisorption can deal with almost all CWAs unlike chemisorption. In this study, we have investigated the inter-pi and inter-dipole-moment interactions between several potential adsorbents and model gases mimicking the CWAs. For the adsorption of non-dipole moment agents, we have observed that the texture profiles (e.g. the uniformity of micropore evolution and specific surface area) play a crucial role in physisorption capacity, but the dipole-non-dipole interaction has a relatively weaker influence on physisorption capacity. In contrast, sulfonated adsorbent, PP-2-S, has boasted superior capacity toward $CH_2O$, meaning that the inter-dipole interaction works well, despite of the relatively inferior texture properties compared to highly activated carbon. According to this insight, we have shed light on porous organic polymers as a potential candidate for chemical weapon adsorbents, in terms of the feasibility of attaching various functional groups. Composites of microporous activated carbons and functionalized porous organic polymers will be studied in future work, thereby paving the way to successfully protect lives from deadly chemical warfare.

## Acknowledgments

This study was funded by center for Research Officers for National Defense, cooperating with the Ministry of National Defense and the Ministry of Science and ICT of the Republic of Korea by Millitech Research Challenge 2022. Tae-hyun Bae and Hyerin Choi at KAIST advised for experiment and analysis.

## Author Contributions

Sanghyeon Park: Conceptualization, formal analysis, investigation, writing, Yuseung Hong: methodology, writing, Hyunseo Choi: methodology, writing, review and editing

## Declarations

### Conflict of Interest

The authors declare that there is no competing financial interests or personal relationships that could have appeared to influence the work reported in this paper.

## References

[1] Kelly Ng, North Korea drops trash balloons on the South. (BBC, 2024) https://www.bbc.com/news/articles/c4nn2p32zrzo, Accessed 29 May 2024

[2] Lim Jong-sun, Kim Cheol-sung, Defense & Technology, 240, 68-7 (1999)

[3] Mingcan Mei, Xutao Hu, Zhen Song, Lifang Chen, Liyuan Deng, Zhiwen Qi, Journal of Molecular Liquids, 348, 118036 (2022)

[4] M. Li., J. Liu., S. Deng., Q. Liu., N. Qi., and Z. Chen., ACS Applied Energy Materials, 4, 7983~7991 (2021)

[5] Yang, Y., Chuah, C. Y., & Bae, T. H, *Chemical Engineering Journal*, *358*, 1227-1234 (2019)

[6] Shaojun Dou, Liang Hao, Hong Liu, Chemical Engineering Science, 262, 117988 (2022)

[7] Bartelt-Hunt, S. L., Knappe, D. R. U., & Barlaz, M. A, *Critical Reviews in Environmental Science and Technology*, *38*(2), 112–136 (2008)

[8] Bardestani R, Patience GS, Kaliaguine S., *Can J Chem Eng.*, 97, 2781–2791 (2019)

[9] Yang, Y., Goh, K., Chuah, C. Y., Karahan, H. E., Birer, Ö., & Bae, T.-H., *Carbon,* *155*, 56-64 (2019)

[10] Lodewyckx, P., Elsevier, 7, 475-528 (2006)

[11] Lee, Y. R., Jang, M. S., Cho, H. Y., Kwon, H. J., Kim, S., & Ahn, W. S., *Chemical Engineering Journal*, *271*, 276-280. (2015)

[12] Xu, Z., Zhang, T., Yuan, Z., Zhang, D., Sun, Z., Huang, Y., ... & Zhou, Y., *RSC advances*, *8*(66), 38081-38090 (2018)

[13] Bedia, J., Peñas-Garzón, M., Gómez-Avilés, A., Rodriguez, J. J., *C*, *6*(2), 21 (2020)

[14] Alothman, Z. A., *Materials*, *5*(12), 2874-2902 (2012)

[15] Sing, K. S., *Pure and applied chemistry*, *57*(4), 603-619 (1985)

[16] Horváth, G., & Kawazoe, K., *Journal of Chemical Engineering of Japan*, *16*(6), 470-475 (1983)

[17] Barrett, E. P., Joyner, L. G., & Halenda, P. P., *Journal of the American Chemical society*, *73*(1), 373-380 (1951)

[18] Brunauer, S., Emmett, P. H., & Teller, E., *Journal of the American chemical society*, *60*(2), 309-319 (1938)

[19] Lu, S., Liu, Q., Han, R., Guo, M., Shi, J., Song, C., ... & Ma, D., *Journal of Environmental Sciences*, *105*, 184-203 (2021)

[20] Kim, C. H., Lee, S. Y., & Park, S. J., *Journal of CO2 Utilization*, *54*, 101770. (2021)

[21] Velasco, L. F., Guillet-Nicolas, R., Dobos, G., Thommes, M., & Lodewyckx, P., *Carbon*, *96*, 753-758. (2016)

[22] Pierce, C., & Smith, R. N., *The Journal of Physical Chemistry*, *54*(6), 784-794. (1950)

[23] Hu, Y., Zhang, L., Wang, Z., Hu, R., & Tang, B. Z., *Polymer Chemistry*, *14*(21), 2617-2623. (2023)